\documentclass{elsart3}
\usepackage{graphicx}

\usepackage{amsmath}
\newcommand{\bea}{\begin{eqnarray}}
\newcommand{\eea}{\end{eqnarray}}
\newcommand{\be}{\begin{equation}}
\newcommand{\ee}{\end{equation}}
\newcommand{\uG}{\underline{G}}

\def\k{{\mathbf k}}
\def\p{{\mathbf p}}
\def\q{{\mathbf q}}
\def\v{{\mathbf v}}

\begin{document}

\begin{frontmatter}

\title{Magnetic field dependence of the
penetration depth of d-wave
superconductors with strong isolated impurities}

\author{\corauthref{cor} Shan-Wen Tsai}
\author{P. J. Hirschfeld}

\address{Department of Physics, University of Florida, Gainesville FL 32611}
\corauth[cor]{Corresponding author. Email: tsai@phys.ufl.edu}

\begin{abstract}
A $d$-wave superconductor with isolated strong non-magnetic impurities
should exhibit an upturn in the penetration depth at low temperatures
\cite{tsai}.  Here we calculate how an external magnetic field
supresses this effect.
\end{abstract}

\end{frontmatter}
In a recent paper \cite{tsai} we presented a calculation of
current response of a $d$-wave superconductor containing a
single impurity and showed that it is singular in the
low-temperature limit, leading in the case of strong scattering to
a $1/T$ term in the penetration depth.
For a small number of such impurities, this low-$T$ upturn could be
observable in cuprate superconductors. An estimate of the size of this
effect and of the
temperature range in which it should occur agrees
with experiments reported by
Bonn {\it et al.} \cite{bonnetal} in which upturns correlated
with disorder introduced by Zn atoms in YBCO have been observed.
Here we consider the influence of a magnetic field on this effect.

Since the low-temperature upturn in the penetration depth is due to
the large number of quasiparticle excitations near the nodal directions,
as is the similar upturn in the case of Andreev surface states
\cite{walter,alff,prozorov,barashetal}, we might expect any physical 
effect which
smears the gap nodes to cut off the upturn. In particular, the orbital
coupling to an applied magnetic field (nonlinear electrodynamics) will
suppress the upturn as it does in the Andreev case. We need to add the
Doppler shift of the quasiparticle energy $i\omega\rightarrow
i\omega +\v_s \cdot \k$ and follow the same steps as in
Ref. \cite{tsai} to calculate the magnetic field dependence of the
upturn. Here $\v_s$ is the local superfluid velocity
and a typical shift $v_s k_{F\perp} \approx (H/H_0) \Delta_0$,
where $H$ is the applied field and $H_0 = 3 \Phi_0/(\pi^2 \xi_0 \lambda_0)$
is of the order of the thermodynamic critical field.

The Nambu propagator for a pure $d$-wave superconductor in the presence of
an external magnetic field is $\uG^0_{\k}(\omega_n)=
[(i\omega_n+\v_s\cdot\k) \tau_0 +\xi_\k \tau_3+\Delta_\k\tau_1]/D_\k$,
where $D_\k\equiv (\omega_n-i\v_s\cdot\k)^2 + \xi_\k^2+\Delta_\k
^2$, $\xi_\k\equiv \epsilon_\k-\mu$, and the $\tau_i$ are Pauli matrices.
In the presence of one single $\delta$-function impurity, the Green's
function can be expressed exactly in terms of the Green's function of the
pure system and the $T$-matrix for the impurity.
The $\tau_0$-component of the $T$-matrix is $T_0(\omega) = (\pi N_0)^{-1}
G_0(\omega)/(c^2-G_0(\omega)^2)$ where $c$ is cotangent of the $s$-wave
scattering phase shift ($c=0$ corresponds to infinitely strong scattering)
and $G_0=(1/2 \pi N_0){\rm Tr}\sum_\k \uG^0_{\k}(\omega)$. Here we consider
the case in which the superfluid velocity is in the $x$-direction, so the
Doppler shift $\v_s\cdot\k_F$ at the four nodes is $\omega_{1,4}=-\omega_{2,3}
=\omega_s=v_sk_F/\sqrt{2}$. The integrated Green's function $G_0(\omega)=
(\frac{i\omega_+}{\pi N_0 \sqrt{\omega_+^2+\Delta_{0}^2}})
{\sf K}(\frac{\Delta_0}{\sqrt{\omega_+^2+\Delta_{0}^2}})+(\frac{i\omega_-}{\pi N_0
\sqrt{\omega_-^2+\Delta_{\k}^2}}){\sf K}(\frac{\Delta_0}{\sqrt{\omega_-^2+
\Delta_{0}^2}})$
where $\omega_{\pm} = \omega \pm i\omega_s$ and ${\sf K}$ is the complete
elliptic integral of the first kind \cite{walker}. We now use Eq. 3 of Ref.
\cite{tsai} to calculate the paramagnetic response $\delta K^{xx}(\p,\q)$ and
from it obtain the change in the penetration depth $\delta\lambda$ following the procedure as in Ref. \cite{tsai}.
Fig. 1 shows the result for a unitary scatterer ($c=0$) and
$\omega_s/\Delta_0 = 0,0.01,0.03,0.05,0.09$. The integral over momentum
is performed for $p,q \sim 1/\lambda \ll k_F$ and nodal expansion of
the order parameter.  The Matsubara sum has been performed numerically.

\begin{figure}[tbp]
\centering
\includegraphics[width=3in,clip]{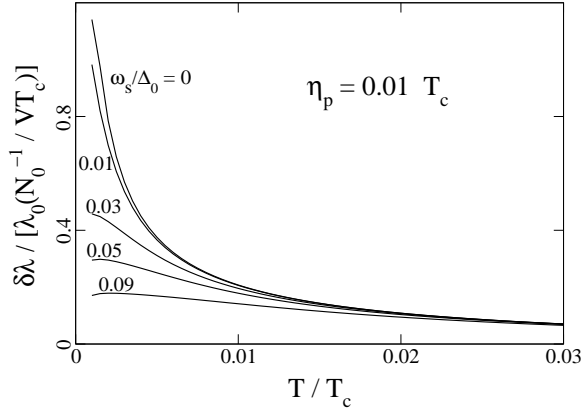}
\caption{Normalized change in the penetration depth due to single isolated
impurity vs. normalized temperature $T/T_c$ for $\omega_s/\Delta_0=0,0.01,0.03,
0.05,0.09$.}
\label{fig1}
\end{figure}

In order for the upturn to be observable, a finite number of
isolated strong scatterers must be present in the sample. An estimate
of the likelihood of occurence of these rare configurations for
randomly distributed impurities in order to extend the theory
to finite disorder can be found in Ref. \cite{tsai}.  In real systems
clustering may take place and we expect
dependence not only on the concentration of impurities but
also on the nature of the statistical distribution of
the impurities in the sample. Indeed measurements on YBCO single
crystals doped with the same amount (0.31\%) of Zn show strong
sample-to-sample dependence \cite{bonnetal}.

In the absence of the external magnetic field, the parameter
$\eta_p = v_F^z/\lambda$ sets a lower bound for the temperature in
which the upturn occurs. A rough estimate for the value of the
external magnetic field in which supression of the upturn should occur
is given by $\omega_s \sim \eta_p$, {\it i.e.},
$\tilde{H}/H_0 \sim \xi_{0z}/\lambda$. Typical values for YBCO
($\xi_{0z} \approx 5\AA$, $\lambda \approx 1500\AA$, $H_0\approx 2.5T$)
give $\tilde{H} \approx 8mT$.

Observation of this effect would represent a remarkable example of
the strong influence that statistically rare impurity
distributions can sometimes have on the macroscopic properties of
a condensed matter system. It can be distinguished from the
similar effect in a $d$-wave superconductor due to Andreev surface
states by its disorder dependence and by its independence of field
orientation and crystal shape.

\end{document}